\documentclass[twocolumn,showpacs,preprintnumbers,amsmath,amssymb,superscriptaddress]{revtex4}
\pdfoutput=1
\usepackage[pdftex]{graphicx}
\usepackage[pdftex,colorlinks=true,urlcolor=blue,linkcolor=black]{hyperref} %online version

\usepackage{graphicx}% Include figure files
\usepackage{dcolumn}% Align table columns on decimal point
\usepackage{bm}% bold math

\begin{document}

\title{Fermionization of Two Distinguishable Fermions} % Force line breaks with \\

%\author{Jochim group}

\author{G.\,Z\"urn$^*$}
\affiliation{Physikalisches Institut, Ruprecht-Karls-Universit\"at Heidelberg, Germany}
\affiliation{Max-Planck-Institut f\" ur Kernphysik, Saupfercheckweg 1, 69117 Heidelberg, Germany}
\email{gerhard.zuern@physi.uni-heidelberg.de}
 \author{F.\,Serwane}
\affiliation{Physikalisches Institut, Ruprecht-Karls-Universit\"at Heidelberg, Germany}
\affiliation{Max-Planck-Institut f\" ur Kernphysik, Saupfercheckweg 1, 69117 Heidelberg, Germany}
\affiliation{ExtreMe Matter Institute EMMI, GSI Helmholtzzentrum f\"ur Schwerionenforschung, Darmstadt, Germany}
 \author{T.\,Lompe}
\affiliation{Physikalisches Institut, Ruprecht-Karls-Universit\"at Heidelberg, Germany}
\affiliation{Max-Planck-Institut f\" ur Kernphysik, Saupfercheckweg 1, 69117 Heidelberg, Germany}
\affiliation{ExtreMe Matter Institute EMMI, GSI Helmholtzzentrum f\"ur Schwerionenforschung, Darmstadt, Germany}
 \author{A.\,N.\,Wenz}
\affiliation{Physikalisches Institut, Ruprecht-Karls-Universit\"at Heidelberg, Germany}
\affiliation{Max-Planck-Institut f\" ur Kernphysik, Saupfercheckweg 1, 69117 Heidelberg, Germany}
\author{M.\,G.\,Ries}
\affiliation{Physikalisches Institut, Ruprecht-Karls-Universit\"at Heidelberg, Germany}
\affiliation{Max-Planck-Institut f\" ur Kernphysik, Saupfercheckweg 1, 69117 Heidelberg, Germany}
\author{J.\,E.\,Bohn}
\affiliation{Physikalisches Institut, Ruprecht-Karls-Universit\"at Heidelberg, Germany}
\affiliation{Max-Planck-Institut f\" ur Kernphysik, Saupfercheckweg 1, 69117 Heidelberg, Germany}
\author{S.\,Jochim}
\affiliation{Physikalisches Institut, Ruprecht-Karls-Universit\"at Heidelberg, Germany}
\affiliation{Max-Planck-Institut f\" ur Kernphysik, Saupfercheckweg 1, 69117 Heidelberg, Germany}
\affiliation{ExtreMe Matter Institute EMMI, GSI Helmholtzzentrum f\"ur Schwerionenforschung, Darmstadt, Germany}

%\author{G.\,Z\"urn$^{1,2}$}
%\email{gerhard.zuern@mpi-hd.mpg.de}
 %\author{F.\,Serwane$^{1,2,3}$}
 %\author{T.\,Lompe$^{1,2,3}$}
 %\author{A.\,N.\,Wenz$^{1,2}$}
%\author{M.\,Ries$^{1,2}$}
%\author{J.\,E.\,Bohn$^{1,2}$}
%\author{S.\,Jochim$^{1,2,3}$}
%\affiliation{$^1$ Physikalisches Institut, Ruprecht-Karls-Universit\"at Heidelberg, Germany}
%\affiliation{$^2$ Max-Planck-Institut f\" ur Kernphysik, Saupfercheckweg 1, 69117 Heidelberg, Germany}
%\affiliation{$^3$ ExtreMe Matter Institute EMMI, GSI Helmholtzzentrum f\"ur Schwerionenforschung, Darmstadt, Germany}
% \author{J}
% \affiliation{Jaffiliation}

%
 %\email{Second.Author@institution.edu}

%\date{\today}% It is always \today, today,
             %  but any date may be explicitly specified

\begin{abstract}
\small We study a system of two distinguishable fermions in a 1D harmonic
potential. This system has the exceptional property that there is an
analytic solution for arbitrary values of the interparticle interaction.
We tune the interaction strength and compare the measured properties of
the system to the theoretical prediction. For diverging interaction
strength, the energy and square modulus of the wave function for two
distinguishable particles are the same as for a system of two
noninteracting identical fermions. This is referred to as
fermionization. We have observed this by directly comparing two
distinguishable fermions with diverging interaction strength with two
identical fermions in the same potential. We observe good agreement
between experiment and theory. By adding more particles our system can
be used as a quantum simulator for more complex systems where no
theoretical solution is available.

\end{abstract}

% PACS, the Physics and Astronomy
                             % Classification Scheme.
%\keywords{Suggested keywords}%Use showkeys class option if keyword
                              %display desired
\maketitle

A powerful tool for solving complex quantum systems is to map their properties onto systems with simpler solutions. 
For interacting bosons in one dimension there is a one-to-one correspondence of the energy and the square modulus of the wave function $|\psi(x_1,...,x_n)|^2$ to a system of identical fermions \cite{Girardeau1960}. As one consequence the local pair correlation $g^{(2)}(0)$ of an interacting 1D Bose gas vanishes for diverging interaction strength just like in a gas of noninteracting identical fermions. Thus, a large decrease of $g^{(2)}(0)$ in a repulsively interacting 1D Bose gas is strong evidence for the existence of fermionization \cite{Weiss2005}. \\
The many-body properties of such 1D bosonic systems have been studied in \cite{Weiss2004, Haller2009}. 
However, the essential property of a such a gas -- namely the fermionization \cite{Girardeau1960, Petrov2000} -- is already present in a system of two interacting particles, regardless of the particles being identical bosons or distinguishable fermions \cite{Girardeau2010}. 
This two-particle problem is of significant interest because it is the main building block of all 1D quantum systems with short-range interactions. It is also one of the few quantum mechanical systems for which an analytic solution exists. 
In contrast to measurements of bulk properties such as compressibility and collective oscillations or measurements of local pair correlations \cite{Weiss2005}, we access the energy and the square modulus of the wave function of the fundamental two-particle system. We directly observe fermionization of two distinguishable fermions by comparing two distinguishable fermions with two identical fermions in the same potential.
In optical lattices the energy of similar two-particle systems has been measured for large but not diverging interaction strength \cite{Esslinger2006, Sengstock2006}.

We realize such a two-particle system with tunable interaction using two fermionic $^6$Li atoms in the ground state of a potential created by an optical dipole trap and a magnetic field gradient [Figs.\,1(a) and 1(b)]. We can prepare this state with a fidelity of $(93\pm 2) \% $ \cite{Serwane2011}. 
The energy of such two particles interacting via contact interaction -- which is fully described by one parameter, the 1D coupling strength $g$ -- can be analytically calculated for a harmonically trapped 1D system \cite{Busch1998, FrankeArnold2003}. 
\begin{figure} [tb]
\centering
	\includegraphics [width= 8cm] {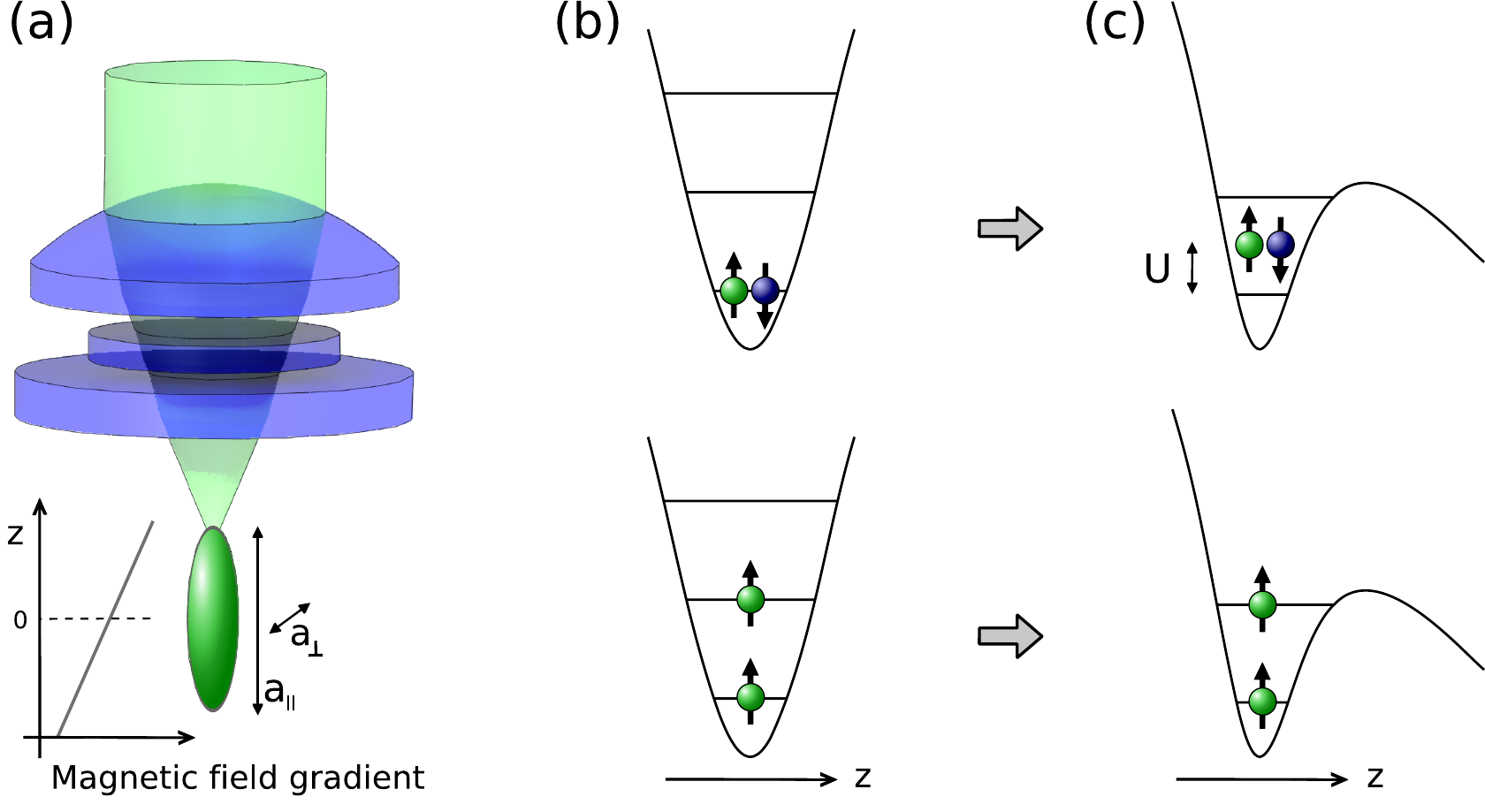}
	\caption{Trap setup and sketch of the performed experiment. (a) Our trap consists of an optical potential created by a tight focus of a laser beam and a magnetic field gradient. (b) Deterministic preparation of two fermions in the ground state of a potential well. (c) We measure the tunneling dynamics through a potential barrier for a repulsively interacting system of two distinguishable fermions for various interaction energies. The mean interaction energy per particle is indicated by the parameter $U$. These results are then compared with the tunneling dynamics of two non-interacting identical fermions in the same potential.}
	\label{fig:figure1}
\end{figure}
The problem can be separated into center-of-mass and relative motion because of the harmonic trapping potential and because the interaction term only depends on the relative distance between the two particles. 
\begin{figure} [tb]
\centering
	\includegraphics [width= 8cm] {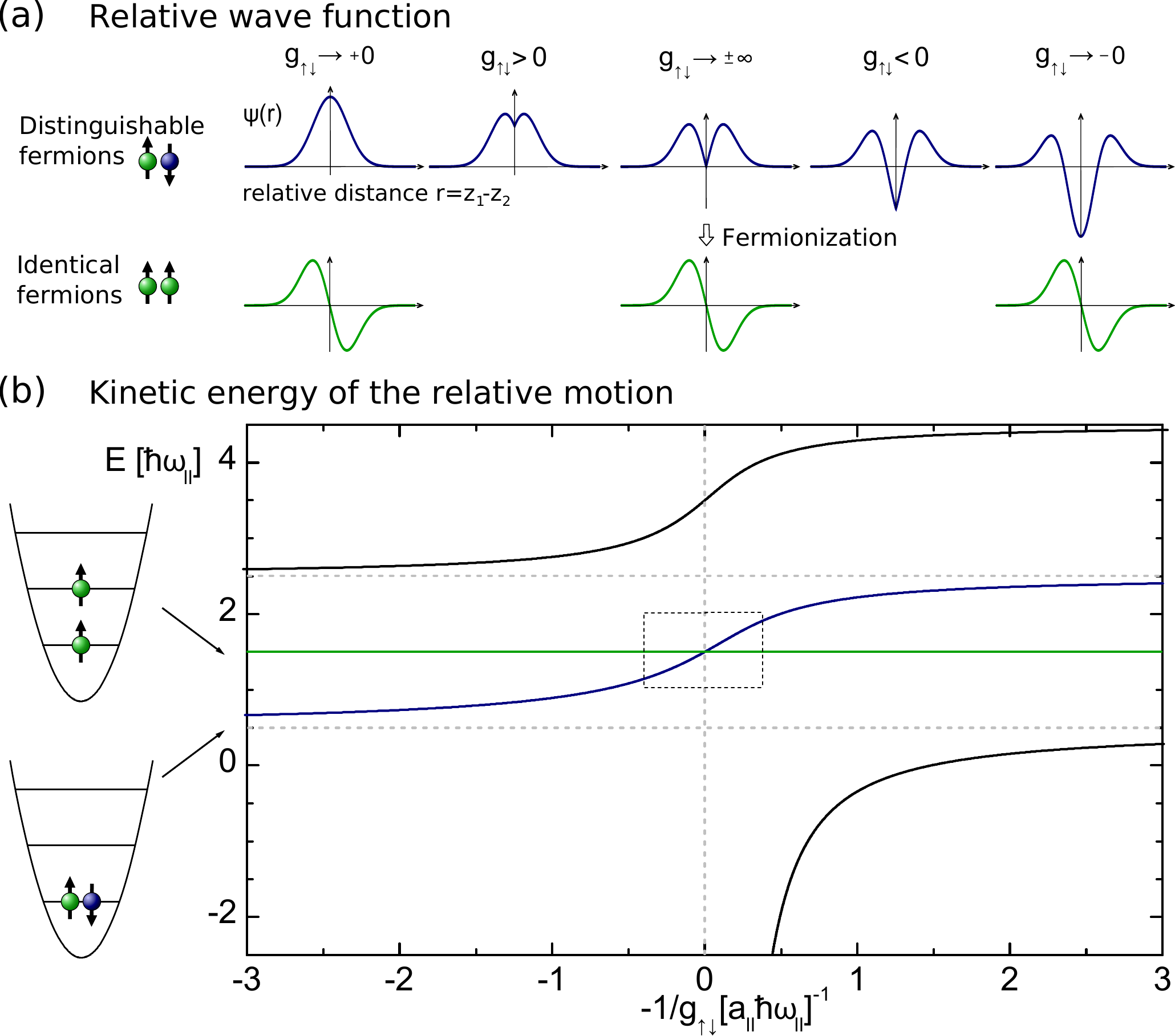}
	\caption{Two particles in a 1D harmonic potential. (a) Relative wave function of two interacting fermions (blue) and two identical fermions (green) in a 1D harmonic potential. For infinitely strong interaction (-1/$g_{\mid\uparrow \downarrow \rangle}\xrightarrow []{}0$) the probability to find the two distinguishable fermions at the same position vanishes. In this case the square modulus of the total wave function of two distinguishable fermions is the same as for two identical fermions. (b) Kinetic energy of the relative motion. The blue and black curves show the energy of two interacting fermions in state $\mid\uparrow \downarrow\rangle$ depending on the coupling strength $g_{\mid\uparrow \downarrow \rangle}$ given in units of $a_{\parallel}=\sqrt{\hbar/ \mu\omega_{\parallel}}$ . 
	The green line shows the energy of two identical fermions in state $\mid \uparrow \uparrow\rangle$. The energy is plotted versus -1/$g_{\mid\uparrow \downarrow \rangle}$ for a better comparison with the experimental results. The experimentally studied region is indicated by the dashed rectangle.}
	\label{fig:figure2}
\end{figure}	
Then the solution can be written as a product of the center-of-mass and the relative wave function. The latter is shown in Fig.\,2(a) for different values of the coupling strength.
For diverging coupling strength the square modulus of the wave function of a system of two distinguishable fermions is the same as for two noninteracting identical fermions. This is the point where fermionization occurs. \\
In our setup the particles are confined in a three-dimensional cigar-shaped potential with an aspect ratio of about 1:10, which can be harmonically approximated with trap frequencies of $\omega_{\parallel}=2 \pi \times (1.234 \pm 0.012)$\,kHz in longitudinal direction and $\omega_{\perp}=2 \pi \times (11.88\pm 0.22)$\,kHz in perpendicular direction. 
It has been shown in \cite{Idziaszek2006} that the energy of two interacting particles in the ground state of such a potential is well described by the 1D solution given in \cite{Busch1998}. Hence we treat our system in this 1D framework.
The combined optical and magnetic potential in one-dimensional form reads: 
\begin{equation}
V_{r=0}(z)= pV_0(1- \frac{1}{1+(z/z_r)^2})- \mu_m \,B' z,
\end{equation}
where $V_0= k_B \,3.326 \,\mu$K  is the initial depth of the optical potential, $p$ is the optical trap depth in units of the initial trap depth, $z_R= \frac{\pi \, w_0^2}{\lambda}$ is the Rayleigh range of the optical trapping beam with minimal waist $w_0= 1.838 \, \mu$m and wavelength $\lambda=1064$ nm, $\mu_m$ is the magnetic moment of the atoms and $B'= 18.92$ G/cm is the strength of the magnetic field gradient. The determination of the trap parameters is described in the Supplemental Material. 
\\
The 1D coupling constant $g$ can be calculated from the 3D scattering length $a_{3D}$ and depends strongly on the confining potential, which is characterized by the harmonic oscillator length $a_{\perp}=\sqrt{\hbar/ \mu \omega_{\perp}}$ \cite{Olshanii1998}, where $\hbar$ is the reduced Planck constant and  $\mu=\frac{m}{2}$ the reduced mass of two $^6$Li atoms with mass $m$.
The coupling constant is given by 
\begin{equation}
g=\frac{2\hbar^2a_{3D}}{\mu a_{\perp}^2}\frac{1}{1-Ca_{3D}/a_{\perp}},
\end{equation}
with $C$$=$$-\zeta(\frac{1}{2})$=$1.46$... and $\zeta$ the Riemann zeta function. 
The value of $g$ can be changed by tuning the 3D scattering length via a magnetic Feshbach resonance \cite{inouye1998, Chin2010}. When $a_{3D}$ approaches the extension of the confining harmonic oscillator potential $a_{\perp}$, a confinement-induced resonance (CIR) occurs for $a_{3D}=a_{\perp}/C$ \cite{Bergeman2003, Haller2010CIR}. Fig.\,4(b) shows $g_{\mid\uparrow \downarrow \rangle}$ for two distinguishable atoms in the two lowest $^6$Li hyperfine states, $\mid $F$=\frac{1}{2}$,$m_F$=-$\frac{1}{2}\rangle$ and $\mid $F$=\frac{1}{2}$,$m_F$=$\frac{1}{2}\rangle$ -- labeled $\mid\uparrow\rangle$ and $\mid \downarrow\rangle$ -- as a function of the magnetic offset field \cite{Bartenstein2005}. For two identical fermions s-wave scattering is forbidden and thus $g_{\mid\uparrow \uparrow\rangle}=0$ for all values of the magnetic offset field.\\
To determine the energy of the two-particle system in state $\mid \uparrow \downarrow \rangle$ we modify the trapping potential such that there is a potential barrier of fixed height through which the particles can tunnel out of the trap (see Fig.\,1(c) and Supplemental Material). In the presence of repulsive interactions the energy of the system is increased according to the blue curve in Fig.\,2(b). This decreases the effective height of the barrier and the particles tunnel faster. We allow the particles to tunnel out of the trap for different durations and record the number of particles remaining in the trap.
By choosing an adequate barrier height we ensure that the time scale for tunneling is smaller than the  lifetime of our samples in the ground state (about $60$\,s).
Additionally obtaining meaningful tunneling time constants requires the timescale of the tunneling to be much larger than the inverse longitudinal trap frequencies of $0.7$\,ms.
By averaging over many experimental realizations we obtain the expectation value of the particle number in the potential for different hold times [Fig.\,(3)]. 
\begin{figure} [tb]
\centering
	\includegraphics [width= 8cm] {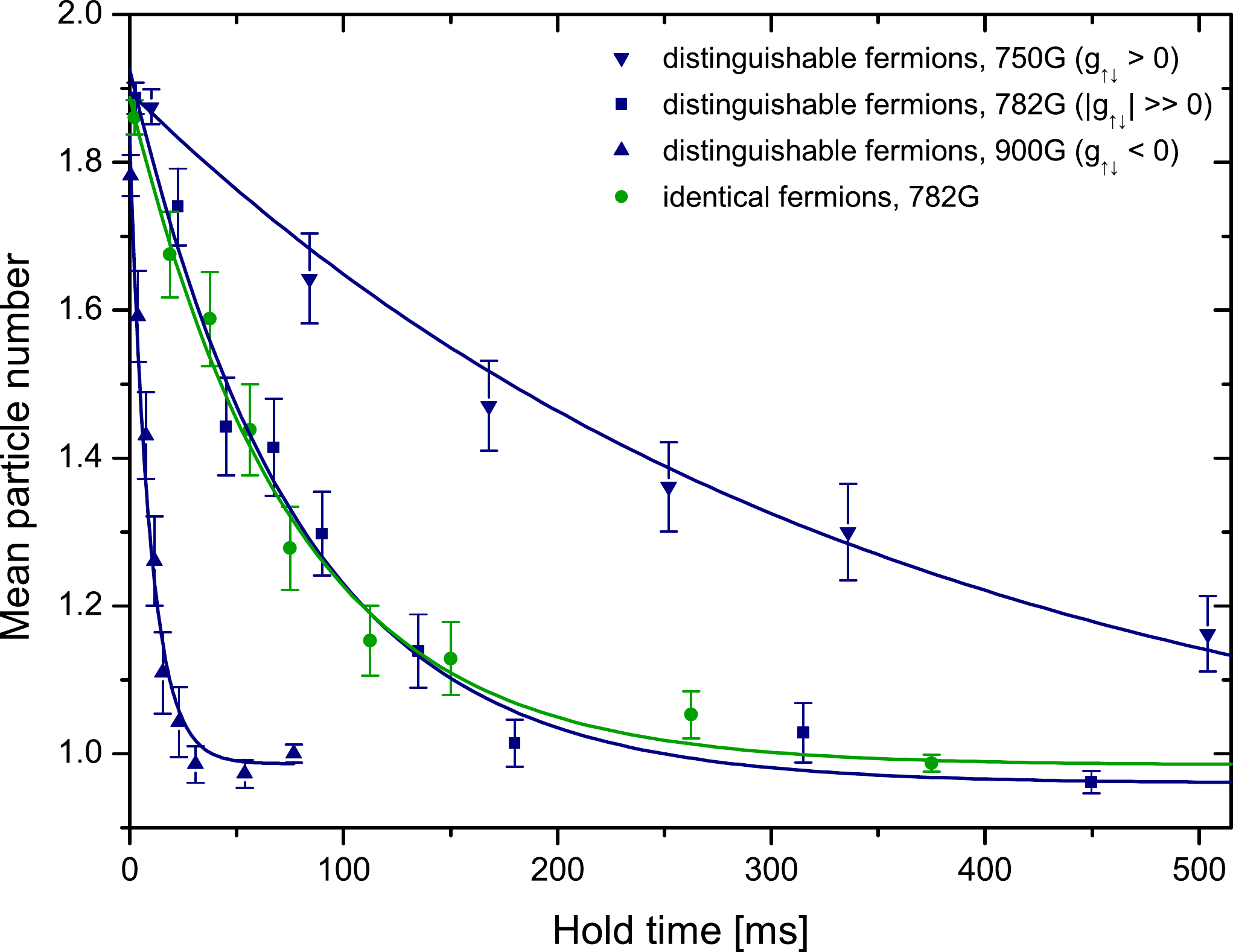}
	\caption{Mean number of particles remaining in the potential well. After modifying the initial potential the particles can tunnel through a barrier of fixed height for a certain hold time. Subsequently, tunneling is switched off and the mean particle number left in the potential is recorded by averaging over many experimental realizations. Exponential fits to the data (solid lines) allow to extract the tunneling time constants of two interacting distinguishable fermions for different interaction strengths (blue) and of two identical fermions (green).	Each data point is the average of about $70$ measurements except for the first and the last data point in each series (about $230$ realizations). The errors are the standard errors of the mean.}
	\label{fig:figure3}
\end{figure}
By performing this measurement for various values of the coupling strength we can determine the dependence of the system's energy on  $g_{\mid\uparrow \downarrow \rangle}$.\\
We find that for the observed range of interaction energies -- which are on the order of $\hbar \omega_{\parallel}$ -- only one particle leaves the potential even for long hold times. In a simple picture this can be explained as follows: If one particle tunnels through the barrier the interaction energy is released as kinetic energy, which leaves the other particle in the unperturbed ground state of the potential. This state has a tunneling time scale much larger than the duration of the experiment.
Thus we can fit exponentials of the form $N(t)=N_{tunnel}\,e^{-\frac{t}{\tau}} + N_{remain}$ to the mean particle number to deduce the tunneling time constant $\tau$ for different magnetic fields. 
The mean numbers of tunneled ($N_{tunnel}$) and remaining particles ($N_{remain}$) are expected to be unity. However, due to the finite preparation fidelity they are slightly lower. In Fig.\,(4) we show the determined tunneling time constants of a system of two interacting fermions for different interaction energies as a function of the magnetic field. 
\begin{figure} [tb]
\centering
	\includegraphics [width= 8cm] {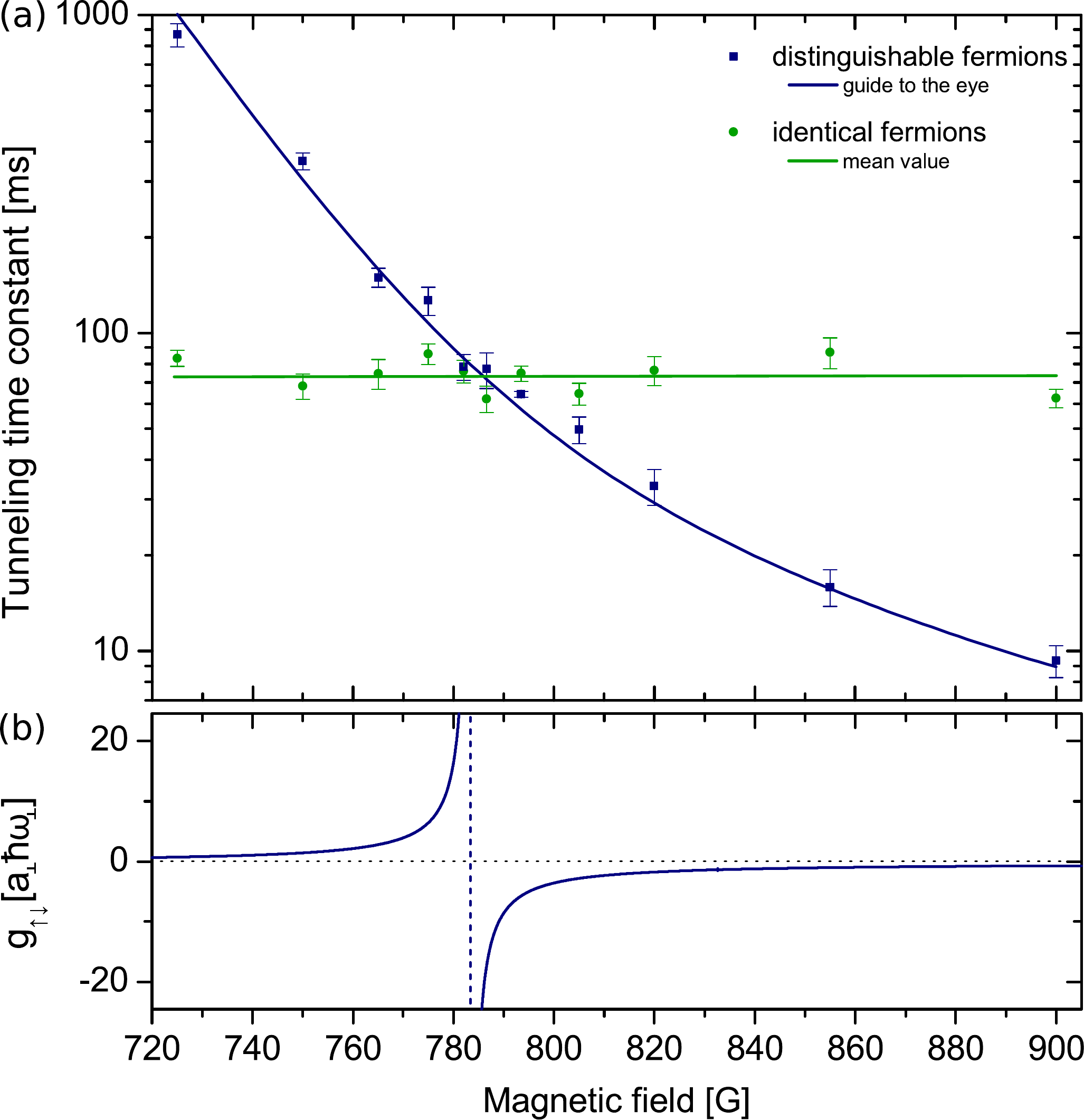}
	\caption{(a) Tunneling time constants for different values of the 1D coupling strength. The tunneling time constant of two repulsively interacting distinguishable fermions (blue curve) decreases by two orders of magnitude with increasing magnetic field. This is attributed to the gain in interaction energy when ramping across the CIR. The tunneling time constant of two noninteracting identical fermions (green line) remains unaffected by the magnetic field within our experimental accuracy. At the magnetic field value where both curves cross we identify the fermionization of two distinguishable fermions. The errors are the statistical errors of the fits shown in Fig.\,3. The blue line is a guide to the eye. (b) One-dimensional coupling constant $g_{\mid\uparrow \downarrow \rangle}$ with a CIR at $(783.4\pm0.4)\,G$. For the calculation we used the perpendicular harmonic oscillator length $a_{\perp}=\sqrt{\hbar/ \mu \omega_{\perp}}$ of the modified potential.}
	\label{fig:figure4}
\end{figure}
We observe a decrease in the tunneling time constant over two orders of magnitude for increasing magnetic field due to the gain in interaction energy caused by the CIR.\\
For a direct comparison of the properties of the two interacting distinguishable fermions with those of two identical fermions we perform the same measurement with two fermions in state $\mid\upuparrows \rangle$ in the same potential [Fig.\,1(c)]. 
The results of these reference measurements are shown in Fig.\,3 and Fig.\,4 (green points).
As the identical fermions are noninteracting we find no dependence of the tunneling time constant on the magnetic field in this measurement.\\
Comparing the results of the two systems we find that the tunneling time constant for the interacting system decreases monotonically with increasing magnetic field and crosses the magnetic field independent tunneling time constant of the two identical fermions. Thus there is one magnetic field value where the tunneling time constants of both systems are equal. At this point both systems must have the same energy. For a 1D system with given energy there is only one unique solution for the square modulus of the wave function. Therefore, right at the observed crossing point of the tunneling time constants the energy and the square modulus of the wave function $|\psi(z_1,z_2)|^2$ of the two interacting distinguishable fermions and the two noninteracting identical fermions must be equal. Hence, exactly at this crossing point the system of two distinguishable fermions is fermionized. As predicted by theory \cite{Busch1998, Girardeau2010} we find the position of the fermionization at the magnetic field value where $g_{\mid\uparrow \downarrow \rangle}$ diverges due to the confinement-induced resonance.\\
For magnetic field values below the CIR we have realized the two-particle limit of a Tonks-Girardeau gas \cite{Girardeau2010}. Above the CIR we have created a super-Tonks state consisting of two particles. The super-Tonks state is a strongly correlated metastable state above the attractive ground state branch [see Fig.\,2(b)]. In a system with particle numbers $\geq3$ inelastic three-body collisions lead to a fast decay of the metastable super-Tonks-Girardeau gas \cite{Haller2009}. In contrast, our two-particle super-Tonks state is stable against collisional losses since there is no third particle available to undergo an inelastic three-body event. \\
To determine the energy of the two interacting fermions from the measured tunneling time constants we use a WKB calculation (see Supplemental Material). This requires knowledge of the potential shape. The parameters of the optical potential are determined by precise measurements of the level spacings in the potential. The final parameter to determine the barrier height is fixed by the measured tunneling time constant of two identical fermions (see Supplemental Material). The energies obtained from the tunneling time constants of two distinguishable fermions are shown in Fig.\,5. \\
\begin{figure} [tb]
\centering
	\includegraphics [width= 8cm] {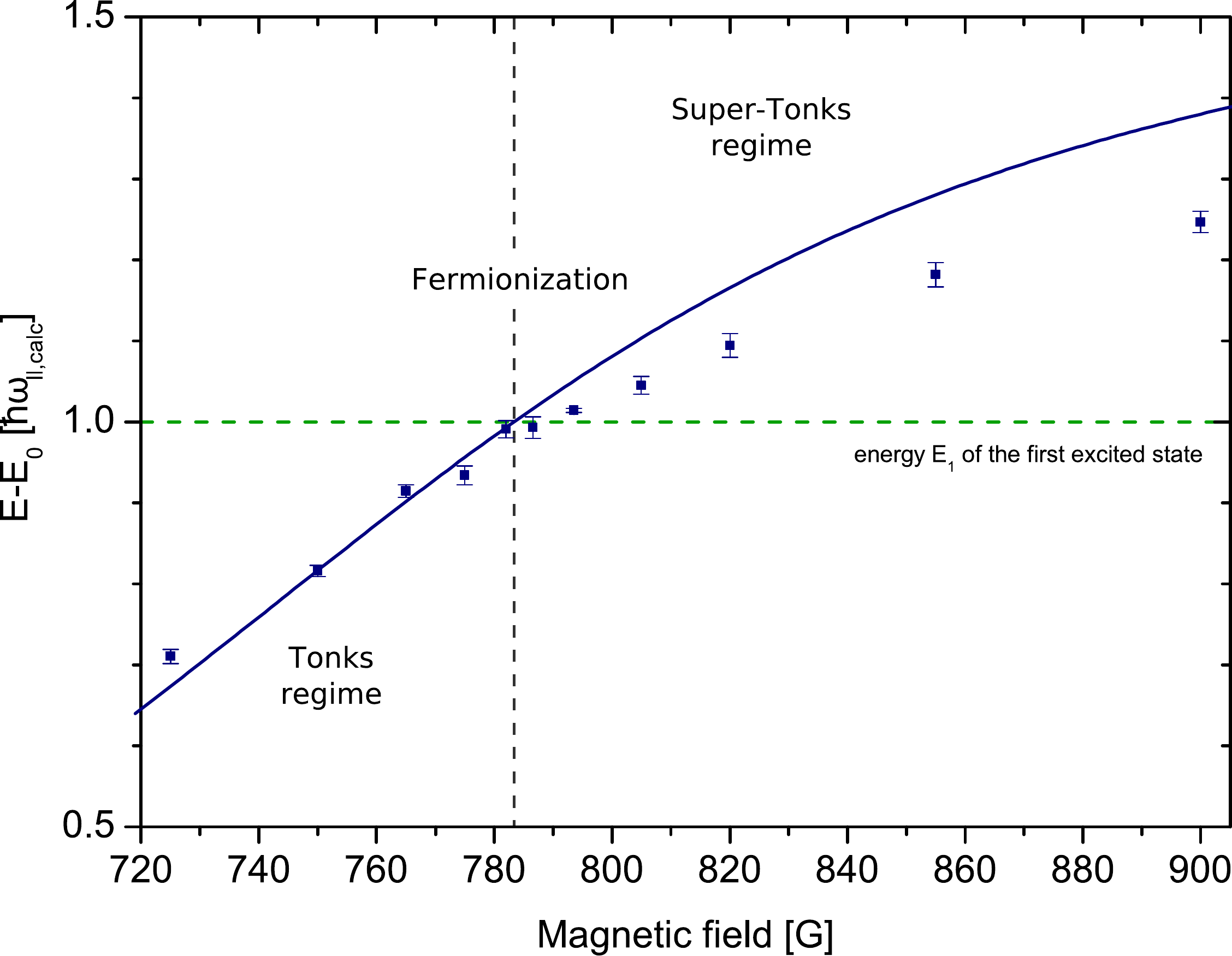}
	\caption{Interaction energy of two fermions for different interaction regimes. By using a WKB based calculation we can determine the energy of two distinguishable fermions at different interaction strengths (blue points) from the tunneling time constants presented in Fig.\,4a. The blue curve shows the expected energy shift for a harmonically trapped system (dashed rectangle in Fig.\,2). }
		\label{fig:figure5}
\end{figure}
We compare these energies to the analytic theory for a harmonic potential \cite{Busch1998} (see Fig.\,2). This theory needs two input parameters, the coupling strength and the level spacing. For the coupling strength we use $g_{\mid\uparrow \downarrow \rangle}$  of our system shown in Fig.\,4(b). For the level spacing we use the energy difference $\hbar\, \omega_{\parallel\,calc}=E_0-E_1=2\pi \hbar \, \times 743$\,Hz between the ground and first excited state of the potential which we calculate using the WKB method. With this approximation the energy obtained from the tunneling measurements and the energy obtained from the analytic theory \cite{Busch1998} are the same at the CIR. 
For the Tonks regime we find excellent agreement of the experimentally determined energy with the theoretical prediction for a harmonic trap. Above the CIR the harmonic theory is not applicable because the second excited state is not bound in our potential. Additionally, we expect deviations for larger energies due to the limited validity of the WKB approximation for energies close to the continuum threshold.
A more precise description could be achieved by adapting the theory described in \cite{Busch1998} to our nonharmonic potential and by using a more accurate theory for the tunneling process \cite{Bardeen1961,Rontani2011} .

In summary, we have measured the interaction energy of two distinguishable fermions as a function of the interaction strength and identified the point of fermionization. The good agreement between our results and theoretical predictions shows that our experiment has the capability to simulate strongly correlated few-body quantum systems. Using the experimental methods established in this work it is straightforward to extend our studies to more complex systems. Simply adding a third particle either in one of the present spin states \cite{Drummond2010_three} or a different spin state \cite{ottenstein2008, Efimov1970} allows us to study a highly nontrivial system where no analytical solution exists. In a few-body system with defined particle number and attractive interaction we could investigate pairing phenomena and thus work towards studying superfluidity in finite systems. 
This has already been investigated in the context of nuclear physics \cite{Migdal1959}.
By dynamically changing the shape of the trapping potential we could simulate a vast amount of different time-dependent quantum systems. A feasible experiment would be to periodically modulate the strength of the magnetic field gradient. This would allow us to study ionizationlike excitations in the strong-field regime \cite{Keldysh1965} which have been studied in ultrafast physics \cite{Corkum1993}. 

We thank H.J. Pirner for valuable theoretical input. We thank T. Busch, M. Rontani and A. Saenz for inspiring discussions and J. Ullrich and his group for their support. This work was supported by the IMPRS-QD (G.Z. and A.N.W.), the Helmholtz Alliance HA216/EMMI, and the Heidelberg Center for Quantum Dynamics.

%\bibliography{zuern}

\section*{SUPPLEMENTAL MATERIAL}
\subsection*{Tunneling potential} 
Our initial samples, which we prepare as described in \cite{Serwane2011}, consist of two noninteracting particles in the ground state of our trap. At the end of the preparation the optical trap depth is $p= 0.795$  (relative error of $p$: $1.3 \times 10^{-3}$) and thus there are 4 bound states in the potential. At this trap depth tunneling from the two lowest levels is completely suppressed. For the measurement of the interaction-induced tunneling we ramp to a trap depth of $p=0.6875$  with a ramp speed of $dp/dt= 0.043 $\, ms$^{-1}$ where tunneling occurs on experimentally accessible time scales. In \cite{Serwane2011} we have estimated the probability of exciting particles when performing a single ramp of the optical potential at this speed to be consistent with $(3\pm1)\%$. To deduce the tunneling time constant we fit the measured particle number with a $\chi$-square minimization algorithm (Levenberg-Marquardt) considering the statistical error of the mean particle number.\\

\subsection*{Magnetic offset field} 
To tune the strength of the interaction we apply magnetic offset fields ranging from $523$\,G to $900$\,G.  At two distinct magnetic field values close to the CIR ($783.2$\,G and $788.9$\,G at $p=0.795$) we observe molecule formation resonances with a width of $0.4$\,G (FWHM). We attribute these resonances to a coupling between the relative motion of the two atoms and the center-of mass motion of a molecular bound state due to the anharmonicity of the trapping potential \cite{Saenz2011, Drummond2011}. We prevent formation of molecules by ramping across those resonances with sufficiently high speed ($20$ G/ms). Thus we can neglect these resonances for our experiments. \\

\subsection*{WKB approximation} 
To determine the energy of the bound states of the potential with the WKB (Wentzel-Kramers-Brillouin) approximation one has to solve the following implicit equation for the energy E:
	\[  \frac{1}{\pi\hbar}\int^{z_b}_{z_a}\,\sqrt{2m(E-V(z))}\, dz=n+\frac{1}{2} \]
with $n$ a positive integer number and $z_a<z_b<z_c$ the solutions of ${V(z)-E=0}$.
The tunneling time constant $\tau$ of a particle with energy E is calculated by $\tau=l\,/ T$
with the transmission coefficient $T=exp\,(-2 \int^{z_c}_{z_b} \sqrt{\frac{2m}{\hbar^2} (V(z)-E)}\, dz$) and the knock-frequency $l=\frac{2\pi\hbar}{E}$. 
For the energy range studied in the experiment we observe that only one of the two particles tunnels through the barrier. This particle has an energy identical to the kinetic energy of the relative motion of the two-particle system. 
Hence we can determine the kinetic energy of the relative motion of the two-particle system by extracting the energy of the tunneled particle from the tunneling time constant $\tau_{exp}$. 
To map $\tau_{exp}$ onto energies we calculate tunneling time constants for a set of energies. By matching these calculated tunneling time constants to the measured ones we determine the kinetic energy of the relative motion of the two-particle systems. A more detailed description of this method will be given in \cite{Serwane2011_thesis}. \\

\subsection*{Determination of the potential shape}
 The potential consists of a cigar-shaped cylindrically symmetric optical potential $V_{opt}$ created by a tightly focused laser beam and a linear magnetic potential $V_{mag}$. The longitudinal part of the potential is given by \nolinebreak{$V_{r=0}(p,z)=V_{opt}(p,z)+V_{mag}(z)=p V_0\,(1-\frac{1}{(1 + (z/z_r)^2))})-\mu_m\,B' z$} where $V_0$ is the initial depth at the center of the optical dipole trap, $p$ the optical trap depth as a fraction of the initial depth, $z_r=\pi\, w_0^2/\lambda$ the Rayleigh range, $\lambda =1064$\,nm the wavelength of the trapping light, $B'$ the magnetic field gradient and $\mu_m$ the magnetic moment of the atoms.\\  
The optical potential is created by a tightly focused laser beam with a power  of $P_0=(265\pm27)\, \mu$W. We assume that the beam has a Gaussian shape, neglecting any aberrations that might be introduced by the optical setup. To deduce the shape of the potential we directly measure the structure of the energy levels in the optical potential by periodically modulating the power or position of the trapping beam. For this we use noninteracting atoms and thus the observed excitation frequencies correspond to the energy differences between the different trap levels. At an optical trap depth of $p=1$ we find level spacings of     
\nolinebreak{$\hbar \omega_{\parallel_{0-1}}=2\pi\hbar \times (1.486\pm0.011)$\,kHz}, \nolinebreak{$\hbar \omega_{\parallel_{0-2}}=2\pi\hbar \times (2.985\pm0.010)$\,kHz} and \nolinebreak{$\hbar \omega_{\parallel_{2-4}}=2\pi\hbar \times (2.897\pm0.020)$ \,kHz}  for the lowest trap states in longitudinal direction where the indices label the energy levels in the trap. 
To obtain the waist we vary $w_0$ and $P_0$ within its error $\sigma_{P_0}$ to minimize $\sum_{(i,j)=(0,1)(0,2)(2,4)}\frac{1}{\sigma_{\omega_{\parallel_{i-j}}}}[(E_{opt\,j}-E_{opt\,i})-\hbar\omega_{\parallel_{i-j}}]^2$  with $E_{opt\,i,j}$ being the energy of the calculated bound states of the varied optical potential. The bound state energies are calculated using a WKB calculation. We find a minimum deviation for $w_0=1.838\mu m$ and $P_0=291.5\,\mu$W resulting in an initial depth at the center of the optical trap of $V_0=k_B \times 3.326$\,$\mu$K. For these parameters two of the three frequencies match the calculated bound states within their errors and all three match within $2 \sigma$.
After fixing the parameters for the optical potential we have to determine the strength of the linear magnetic potential which is created by a magnetic field gradient $B'$. From a levitation measurement we obtain $B'=(18.9\pm0.2)\,$\,G/cm. 
To obtain a more accurate parametrization of the potential barrier we use the tunneling measurement of two identical fermions at a trap depth of $p=0.6875$  as a calibration. We perform a WKB calculation of the tunneling rate and modify the value of $B'$ in the calculation such that the resulting tunneling time constant agrees with the experimentally observed tunneling time constant of $(74.1\pm 2.7)$\,ms. From this we obtain $B'=18.92 $\,G/cm. 
Using this magnetic field gradient the calculation is consistent with the deterministic preparation \cite{Serwane2011} of (2, 4, 6, 8, 10) particles at $p$=(0.6575, 0.7025, 0.7475, 0.7863, 0.8200) in the experiment.\\
We measure the perpendicular trap frequency by exciting noninteracting particles in the pure optical potential $V_{z=0}(r,p)=p V_0\,(1- e^{-\frac{2r^2}{w_0^2}})$. In perpendicular direction we observe two resonances at $2 \omega_x= 2 \pi \times 29.07$\,kHz and $2 \omega_y= 2 \pi \times 28.26 $\,kHz for $p=1$ which we attribute to a slight ellipticity of the trap. Following \cite{Drummond2010, Saenz2011, Drummond2011}  we only consider one CIR which we calculate from the mean perpendicular frequency of $2 \pi \times (14.33\pm0.26)$\,kHz.\\
At $p=0.6875$, where we perform the tunneling measurements, the trap frequency of the perpendicular confinement has to be scaled by $\sqrt{p}$ and is given by $\omega_{\perp}=2 \pi \times (11.88\pm 0.22)$\,kHz. At this depth the mean longitudinal trap frequency of the pure optical potential is given by $\omega_{\parallel}=2 \pi \times (1.234 \pm 0.012)$\,kHz which is calculated from the excitation frequencies $\omega_{\parallel_{0-1}}$ and $\omega_{\parallel_{0-2}}$.\\

\end{document}